\documentclass[11pt,a4paper,onecolumn]{article}
\usepackage[utf8]{inputenc}
\usepackage[english]{babel}
\usepackage{amsmath}
\usepackage{amsfonts}
\usepackage{mathabx}
\usepackage{amssymb}
\usepackage{graphicx}
\usepackage{nth}
\usepackage{stackrel}
\usepackage{multirow}
\usepackage{color}
\usepackage{colortbl}
\usepackage{tikz,tkz-tab}
\usepackage{xspace}

\usepackage[left=2cm,right=2cm,top=2cm,bottom=2cm]{geometry}

\DeclareMathSymbol{\Tau}{\mathalpha}{operators}{"54}
\newcommand{\Ratio}{h}

\newcommand{\Pressure}{p}
\newcommand{\PressureDrop}{q}

\newcommand{\ExtPressure}{P_{tissue}}
\newcommand{\ShearStress}{\sigma}
\newcommand{\Resistance}{\mathcal{R}}
\newcommand{\ResistPoiseuille}{\mathcal{R}_{P}}
\newcommand{\ResistMod}{Z}
\newcommand{\Radius}{R}
\newcommand{\NormSection}{\alpha}

\newcommand{\Length}{L}
\newcommand{\Volume}{V}
\newcommand{\Surface}{S}
\newcommand{\Wall}{\Surface_{\rm{Wall}}}
\newcommand{\Inlet}{\Surface_{\rm{In}}}
\newcommand{\Outlet}{\Surface_{\rm{Out}}}

\newcommand{\Velocity}{v}
\newcommand{\Viscosity}{\mu}
\newcommand{\Density}{\rho}
\newcommand{\Flow}{\Phi}
\newcommand{\Reynolds}{\mathcal{R}e}
\newcommand{\R}{\mathcal{R}e}
\newcommand{\hc}{\left(\frac12\right)^{\frac13}}
\newcommand{\hci}{\left(\frac12\right)^{\frac12}}
\newcommand{\dd}{\mathrm{d\/}}

\newcommand{\Bronchus}{b}
\DeclareMathOperator{\Lambert}{\alpha_{Lambert}}

\newcommand{\TransPressure}{P}
\newcommand{\ith}{\textsuperscript{th}\ }
\newcommand{\cmHHO}{cmH\textsubscript{2}O\xspace}
\newcommand{\LperS}{L.s\textsuperscript{-1}\xspace}

\newcommand{\cred}{\color{black}}

\newcommand{\cb}{\color{black}}

\author{Jonathan Stéphano$^1$, Benjamin Mauroy$^{1,*}$\\
\small $^1${\it Université Côte d'Azur, CNRS, LJAD, VADER Center}\\
\small $^*${\it Corresponding author, {\rm benjamin.mauroy@univ-cotedazur.fr}}}
\title{Wall shear stress distribution in a compliant airway tree}

\date{15 March 2021}

\begin{document}
\maketitle

\begin{abstract}
\cred
The airflow in the bronchi applies a shear stress on the bronchial mucus, which can move the mucus.
The air--mucus interaction plays an important role in cough and in chest physiotherapy (CP).
The conditions under which it induces a displacement of the mucus are still unclear.
Yet, the air--mucus interaction justifies common technics of CP used to help the draining of the mucus in prevalent diseases.
Hence, the determination of the distribution of the shear stress in the lung is crucial for understanding the effects of these therapies and, potentially, improve their efficiency.

We develop a mathematical model to study the distribution of the wall shear stress (WSS) induced by an air flow exiting an airway tree.
This model accounts for the main physical processes that determines the WSS, more particularly the compliance of the airways, the air inertia and the tree structure.

We show that the WSS distribution in the tree depends on the dynamics of the airways deformation and on the air inertia.
The WSS distribution in the tree exhibits a maximum whose amplitude and location depend on the amount of air flow and on the "tissue" pressure surrounding the airways.
To characterize the behavior of the WSS at the tree bifurcations, we derive new analytical criteria related to the airway size reduction in the bifurcations. 

Our results suggest that a tuning of the airflow and of the tissue pressure during a CP maneuver might allow to control, at least partially, the air--mucus interaction in the lung.
\cb
\end{abstract}

\twocolumn
\section{Introduction}

\cred
The lung forms an interface between the ambiant air and the blood.
It is an easy entry point in the organism for toxic or infectious particles.
As a protection, the walls of the bronchi are covered by a mucus layer.
The mucus captures the inhaled particles and is incessantly moved toward the oesopahryngeal region by the mucociliary clearance \cite{bustamante-marin_cilia_2017}.
Once in the oesopharyngeal region, the mucus is either swallowed or expelled by coughing.
In the non-mature or pathological lung, the mucociliary clearance might not be able to drain the mucus correctly and the lung relies on the cough to drain the mucus~\cite{ho_effect_2001, baby_effect_2014}.
The mucus is a complex viscoelastic material that moves as a fluid only when its inner stress overcomes its yield stress \cite{lai_2009}.
During lung's ventilation~\cite{stephano_modeling_2019}, during cough \cite{king_role_1987} or during chest physiotherapy~\cite{mauroy_2011, mauroy_2015, stephano_modeling_2019}, the airflow interacts with the mucus and applies a shear stress that can potentially overcome the mucus yield stress.
Chest physiotherapy is commonly used to compensate the dysfunctions of the mucocilliary clearance and of the cough occurring in prevalent diseases such as asthma, bronchiolitis, COPD or cystic fibrosis \cite{pryor_physiotherapy_1999}.
However, the distribution of the shear stress and the region in the lung where the mucus might be liquified are not well identified as of today.
Hence, we propose in this work to model the airflows in the bronchial tree to study the distribution of the wall shear stress.

Past studies have developed models of the air fluid dynamics in the lungs.
However, the wide range of scales covered by the airways makes lung's modelling a challenge \cite{crampin_computational_2004, longest_use_2019}.
There are two main approaches.
The first approach is to model the fluid dynamics in a subpart of the lung using 3D geometries, either idealised or reconstructed from CT-scans \cite{mauroy_interplay_2003, mauroy_3D_2005, kabilan_characteristics_2007, moghadas_numerical_2011, paz_glottis_2017, farnoud_large_2020, ren_numerical_2020}.
Some studies go further and model the air--mucus interaction in 3D geometries using the volume of fluids method \cite{kumar_numerical_2017, rajendran_mucus_2019} or the thin layer theory \cite{paz_glottis_2017, ren_numerical_2020}.
Such studies are mainly focused on the upper respiratory region and assume the geometries to be rigid.
Micro scale models have also been developed to study mucociliary clearance \cite{lubkin_viscoelastic_2006, jayathilake_three-dimensional_2012} or the respiratory gaz flows in the deep lung (acini) \cite{felici_diffusional_2005, swan_evidence_2011}. 
The models are analysed with high-end computational fluid dynamics based on finite elements or finite volumes, which are not well adapted to large change of scales. 
In such models, good boundary conditions to mimic the response of the regions non accounted for are not trivially determined.
Nevertheless, the resulting predictions are often very detailed and give rich insights on the local dynamics.
The second approach, which we will use in this work, is based on idealised representations of the lungs that reduce the behavior of groups of airways to a single behavior. 
Typically the airways are regrouped by generations, i.e. the number of bifurcations between the airway considered and the root of the tree   \cite{mauroy_2004, mauroy_2011, mauroy_2015, stephano_modeling_2019}.
These approaches allow to mimic the whole airway tree and have led to interesting fundamental insights on the lung \cite{lambert_1982, mauroy_2004, mauroy_2011, mauroy_2015, noel_interplay_2019, noel_origin_2020}. 

However, in either approach, no model did include both the role of the air inertia and of the airways compliance, although these are known to affect notably the flow properties in the lung \cite{mauroy_interplay_2003, mauroy_2015}.
Indeed, the airways diameters depend on the airways inner air pressure, the airways inner air pressure depends on the air fluid dynamics, potentially affected by inertia, and the air fluid dynamics depends on the airways diameters.
These interactions occur in a tree-like structure which increases the complexity of the system \cite{mauroy_2004}.
\begin{figure}[t!]
\centering
\includegraphics[width=6cm]{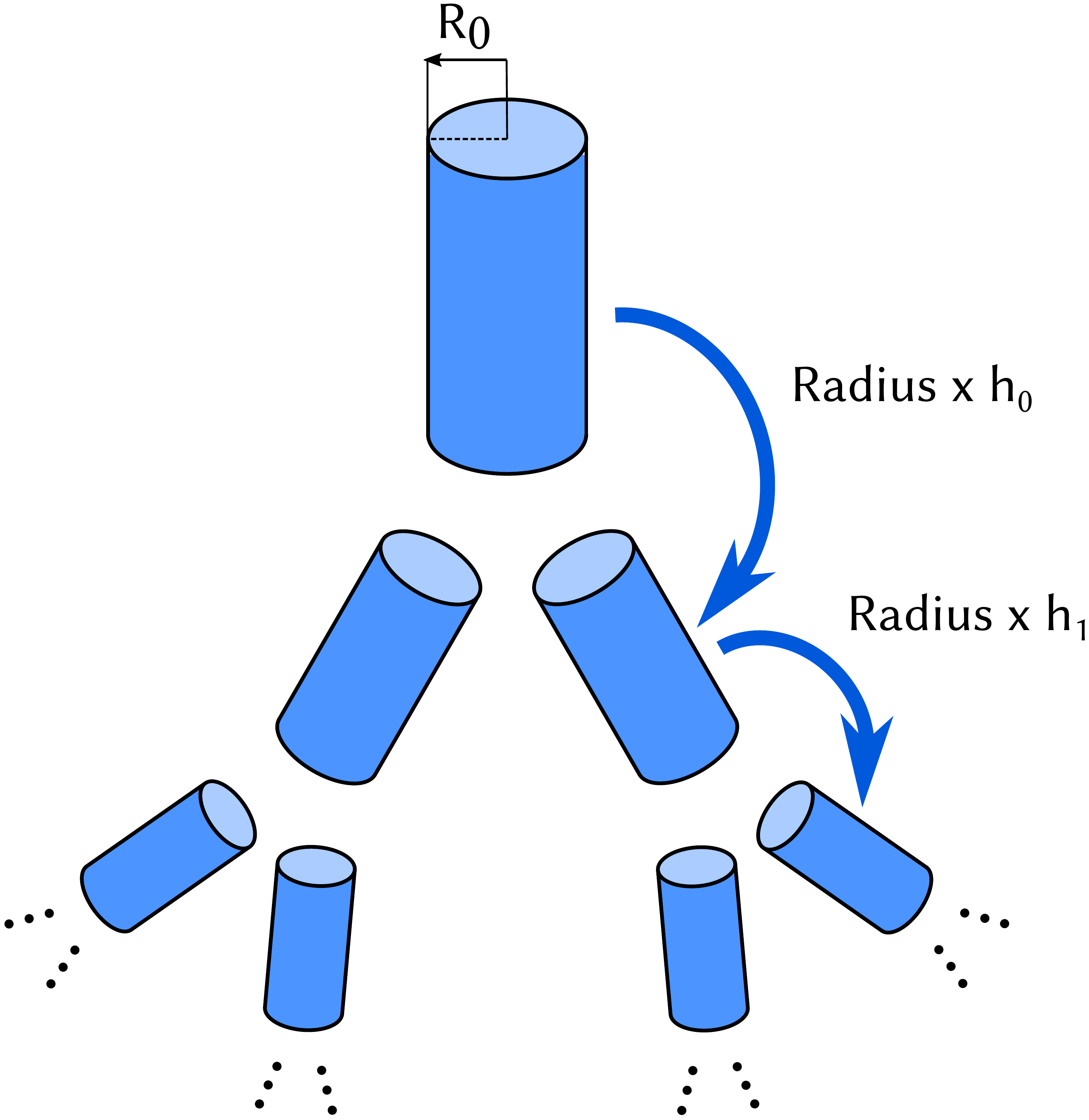}
\caption{Schematics of the airway tree structure. 
The number on the airways corresponds to the generation index of that airway. The generation index corresponds to the number of airways between the airway considered and the root of the tree. 
From the generation $i$ to the generation $i+1$, the diameter of the airways is multiplied by a ratio $\Ratio_i < 1$.}
\end{figure}
In this framework, we propose an original model to study for the first time the physics of an inertial flow in an airway tree with compliant airways. 
Because the mucus thickness is far smaller than the airway diameters \cite{karamaoun_new_2018}, we neglect the mucus layer and assume that the shear stress induced by the air on the airways wall is a good estimation of the shear stress felt by the mucus in the lung.

For the first time, a study is able to propose a detailed physical analysis of the interactions between the airflows and the deformation of the airways in the tree structure.
Our model predicts the existence of a maximum of wall shear stress (WSS) within the tree.
We uncover a new analytical criterion that allows to explain the existence of this maximum.
This criterion is based on the physics occurring in the bifurcations and allows to determine whether the shear stress will increase or decrease when the air is going through a bifurcation.
The criterion depends on the geometries of the deformed airways and on the amount of air inertia.
We show that the location of the maximum of WSS is a function of the total airflow rate in the tree and of an homogeneous "tissue pressure" that surrounds the airways.
Modulating these two quantities allows, to some extent, to control the location and amplitude of the maximum of wall shear stress.

This model can be considered as an idealised representation of the lung, with the tissue pressure being a representation of the pressure in the lung's parenchyma.
In that framework and in the limit of our hypotheses, our model brings for the first time a better understanding of the shear stress induced by the air on the mucus in the lung.
Moreover, our work highlights the underlying biomechanics involved in high ventilation regimes and in some of the most common chest physiotherapy technics.
This study might pave the way for future improvements of the manipulations of chest physiotherapy.
\cb

\section{Model}
\label{model}

The airflow inside the airway induces an air pressure that opposes the mechanical pressure that surrounds the airways and that we will call from now on the tissue pressure. 
The balance between these two pressures affects the diameters of the deformable airways, which in turn affects the airflow inside the airway. 

\subsection{Model of the bronchial tree}

We mimic the bronchial tree using the model of the human lung from \cite{lambert_1982}. 
The bronchial tree is represented by a cascade of bifurcating cylinders with radius $R$ and length $L$ representing the bronchial airways. 
The generation of an airway is an index that counts the number of bifurcations on the path between the root of the tree, that mimics the trachea (index $0$), and the airway under consideration.
Each cylinder has a fixed length and a variable diameter that is adjusted depending on the transmural pressure that it is submitted to. 
The transmural pressure of an airway is the difference between its inner air pressure and the pressure surrounding the airways, the tissue pressure. 
In our model, the tissue pressure is assumed homogeneous and the air pressure results from the air flow inside the airway. 
The bifurcations are assumed symmetric, consequently all the cylinders with the same generation index have exactly the same physical and geometrical properties. 
Hence, it is sufficient to study one single airway per generation. 
The ratio between the diameter of a cylinder in the generation $i+1$ and a cylinder in the generation $i$ is called the reduction ratio and is denoted $\Ratio_i$.

In \cite{lambert_1982}, the authors propose a model for static compliant airways based on bronchi data. 
In a generation with index $i$, this model links a normalized section of the airways $\NormSection_i$ to the transmural pressure $\TransPressure_i$.
The section is normalized relatively to a maximal surface area $S_{max,i}$ reached for an infinite transmural pressure.
Since we hypothesize that the airways remain cylindrical whatever the transmural pressure, the model can be expressed in terms of the airways radii.
Hence, the maximal surface area $S_{max,i}$ corresponds to a maximal radius $R_{max,i}$ and $S_{max,i} = \pi R_{max,i}^2$. 
Finally, the radius $R_i$ of an airway in the generation $i$ is related to the normalised surface area of the airway by $R_i = R_{max,i} \sqrt{\NormSection_i}$.
The reformulated Lambert's model writes
\begin{equation}
\label{eq:Lambert}
\left\{
\begin{array}{lcl}
\NormSection_i & \stackrel[\TransPressure_i\le 0]{}{=}& \NormSection_{0,i}\ (1-\TransPressure_i/P^-_i)^{-n^-_i}\\
\NormSection_i & \stackrel[\TransPressure_i\ge 0]{}{=}& 1 - (1-\NormSection_{0,i})\ (1+\TransPressure_i/P^+_i)^{-n^+_i}\\
R_i &=& R_{max,i} \sqrt{\NormSection_i}
\end{array}
\right.
\end{equation}
The quantities $\NormSection_{0,i}$, $P^\pm_i$, $n^\pm_i$ and the maximal possible radius for the airway in generation $i$, $R_{max,i}$, are positive fixed data given by the model.
They depend on the generation index $i$ only.\\
\begin{figure}
\includegraphics[width=\linewidth]{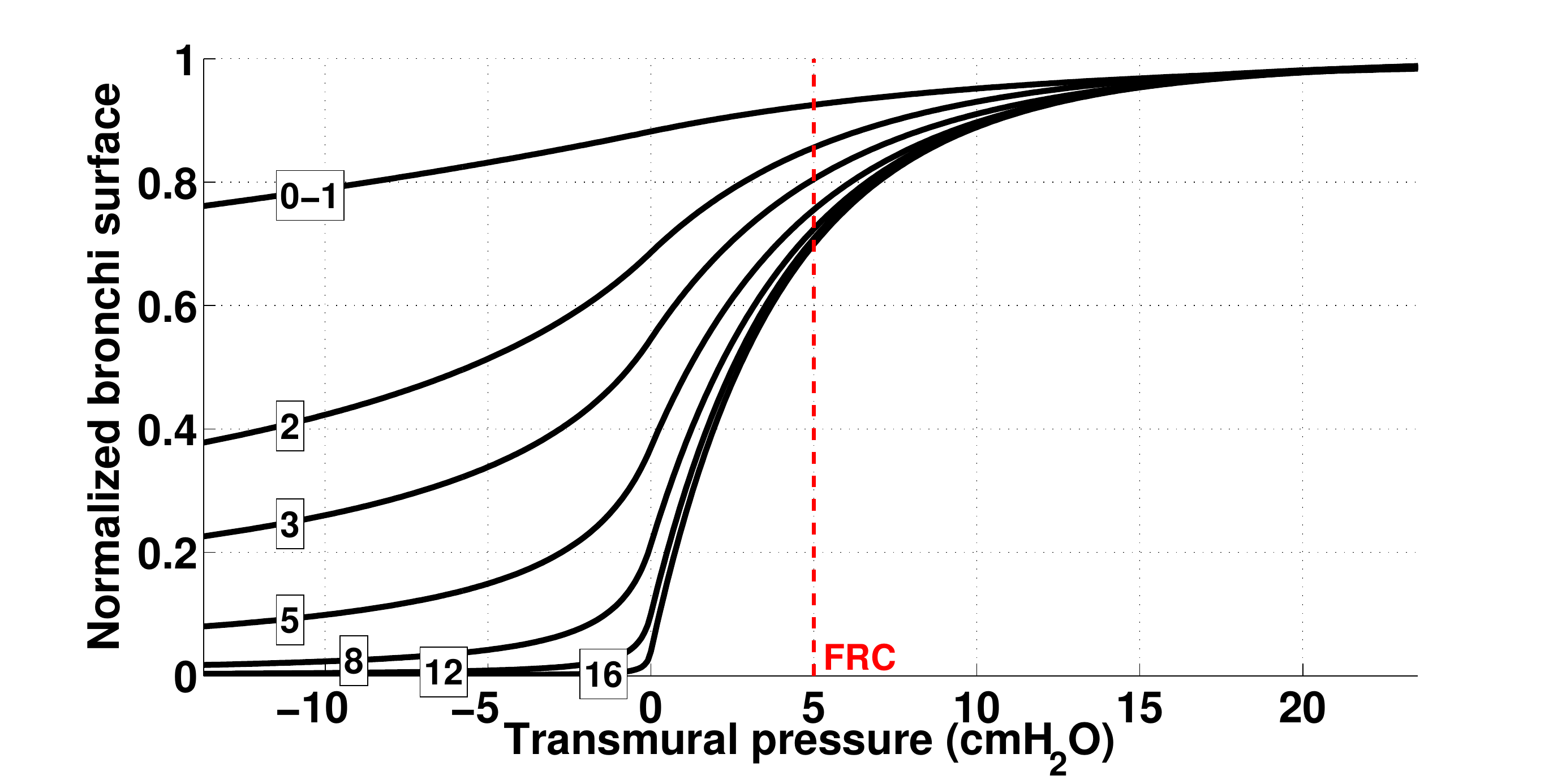}
\caption{Model of compliant airways from \cite{lambert_1982}. The curves represent the normalised surface areas of the airways of a same generation as a function of the transmural pressure in these airways. 
The normalisation is made relatively to a maximal possible surface area of the airways.
The transmural pressure is the difference between the inner (air) and outer pressures (surrounding tissue) of the airways. 
The boxed index indicates the generation of the corresponding airways.
}
\label{lambertFig}
\end{figure}
Since an healthy mucus layer is very thin, from about $3$ micrometers to $10$ micrometers \cite{karamaoun_new_2018}, its influence on the airways radii and on the air fluid dynamics can be neglected.

\subsection{Fluid dynamics in the airways}

We consider a cylindrical airway with a radius $\Radius$ and a length $\Length$.
We will assume the velocity profile to be stationary, axisymmetric, invariant along the airway axis and parallel to the airway axis.
A volumetric integration of the Navier-Stokes equations with the above hypotheses leads to a relationship between the wall shear stress $\ShearStress$ and the pressure drop $q$ through the airway,
\begin{equation}
\label{eq:shear2pressure}
\ShearStress = -\PressureDrop \frac{\Radius}{2\Length}
\end{equation}
Details of the computation are given in the Supplementary Materials (\ref{sigmaq}).
Since we are studying the outgoing flow that is positive, the pressure drop is negative.
Hence, the wall shear stress is positive.

\begin{table*}
\centering
\begin{tabular}{|c|c|c|c|c|}
\hline
\multirow{2}{*}{Generation}  & \multicolumn{2}{c}{FRC} & \multicolumn{2}{|c|}{Maximal dilation}\\
\cline{2-5}
 & Radius (mm) & Ratio & Radius (mm) & Ratio\\
\hline
1 & 8.18 & 0.708 & 8.69 & 0.707\\
\rowcolor[gray]{.9}
2 & 5.79 & 0.687 & 6.14 & 0.769\\
3 & 3.98 & 0.716 & 4.72 & 0.790\\
\rowcolor[gray]{.9}
4 & 2.85 & 0.740 & 3.73 & 0.802\\
5 & 2.11 & 0.711 & 2.99 & 0.769\\
\rowcolor[gray]{.9}
6 & 1.50 & 0.733 & 2.30 & 0.783\\
7 & 1.10 & 0.736 & 1.80 & 0.783\\
\rowcolor[gray]{.9}
8 & 0.810 & 0.756 & 1.41 & 0.801\\
9 & 0.612 & 0.755 & 1.13 & 0.787\\
\rowcolor[gray]{.9}
10 & 0.462 & 0.751 & 0.889 & 0.792\\
11 & 0.347 & 0.769 & 0.704 & 0.805\\
\rowcolor[gray]{.9}
12 & 0.267 & 0.798 & 0.567 & 0.834\\
13 & 0.213 & 0.831 & 0.473 & 0.879\\
\rowcolor[gray]{.9}
14 & 0.177 & 0.847 & 0.416 & 0.882\\
15 & 0.150 & 0.873 & 0.367 & 0.902\\
\rowcolor[gray]{.9}
16 & 0.131 & 0.878 & 0.331 & 0.894\\
17 & 0.115 &  & 0.296 &  \\
\hline
\end{tabular}
\caption{Radii and radii reduction ratios in the airway tree as a function of the generation index, from \cite{lambert_1982}. Two states are presented: {functional residual volume (FRC)} (no airflow, transmural pressure of $5$ \cmHHO) and {maximal dilation} (no air flow, infinite transmural pressure).}
\label{ratioLambert}
\end{table*}

For each airway, the pressure drop $\PressureDrop$ between its two extremities is linked to the flow rate $\Flow$ going through the airway and to the hydrodynamic resistance $\Resistance$ of the airway,
\begin{equation}
\PressureDrop = \Resistance\ \Flow
\label{eq:PressureFlow}
\end{equation}
The resistance depends on the airway's geometry and on the air fluid dynamics. 
In the case of a Poiseuille flow -i.e. slow, fully developed regime-, the hydrodynamic resistance $\ResistPoiseuille$ depends only on the airway geometry,  
\begin{equation}
\ResistPoiseuille = \frac{8\Viscosity \Length}{\pi\Radius^4}
\label{eq:Poiseuille}
\end{equation}
The quantity $\Viscosity$ is the viscosity of the air. 
However, in higher regimes, the air flow is affected by the inertial effects and the hydrodynamic resistance becomes larger than in Poiseuille's regime, $\Resistance = \ResistMod\ \ResistPoiseuille$ where the prefactor $Z$ is larger than $1$.
Various models have been proposed to express the prefactor $\ResistMod$ \cite{pedley_1970, Florens_2011}.
We choose to use the velocity profile proposed in \cite{tawhai_2004}. 
It accounts for inertia using a power law.
The axial velocity depends on the position $r$ on the radius of the airway, 
\begin{equation*}
v_z(r) = v_0 \left( 1 - \left(\frac{r}{R}\right)^{\alpha}\right) \text{ with $\alpha = \max \left(2, \frac{\Reynolds}{150}\right)$}
\end{equation*}
where $\Reynolds$ is the Reynolds number $\Reynolds = \frac{4\Density \Flow}{\Viscosity\pi\Radius}$. 
Then, we can compute the flow rate $\Phi$ through the airway, $\Phi = \frac{\alpha}{\alpha+2} \pi v_0 R^2$, and using the relationship between the wall shear stress $\sigma = \mu \frac{d v_z}{d r}(R) = -\mu \alpha v_0 / R$ and the pressure drop $q$, $\sigma = -q R / (2L)$, we can deduce that $q = \frac{\alpha + 2}{4} \ResistPoiseuille \Phi$.
Hence, the expression of $\ResistMod$ in our model is 
\begin{equation*}
\ResistMod = \max\left(1,\frac12 + \frac{\Reynolds}{600}\right)
\end{equation*} 
Consequently, the resistance factor $\ResistMod$ has a lower bound equal to $1$ that is reached for $\Reynolds = 300$ and that corresponds to the Poiseuille's regime.

Finally, we set the density of the air to $\Density = 1.225$ kg.m$^{-3}$, and its viscosity to $\Viscosity = 1.70 \ 10^{-5}$ Pa.s. 
The reference pressure is set to $0$ at the opening of the first generation, that corresponds to the trachea.
Notice that, with this hypothesis, we do not account for the nasopharyngeal or buccopharyngeal pathway to the air pressure.
Actually, the error made with this hypothesis may be of various importance, depending on the flow rate and on the sizes of the airways.

\subsection{Resolution of the model's equations and validation.}
The global model leads to a set of non linear equations for each generation of the airway tree.
The generations are coupled by the conservation of the air flow and by the continuity of the air pressure.
The equations are solved using a Newton algorithm, implemented in C++ with the library Eigen \cite{eigenweb}.
The set of equations, its properties and the numerical process are detailed in the Supplementary Materials (\ref{system}, \ref{properties} and \ref{numerics}).

\cred
To validate our model, we ran it with simple configurations whose behaviors were either straightforward or predicted by the analytical results presented in the section \ref{analytic}. 
Hence, we tested trees with fractal geometries and rigid airways as in \cite{mauroy_2004}, trees at different scales reduced to a single deformable airway or to a single bifurcation.
\cb

\section{Links between the tree geometry and the wall shear stress distribution}

\begin{table*}
\centering
\begin{tikzpicture}
\tkzTabInit[lgt=7.5, espcl=1.5, deltacl=.25]{reduction ratio $\Ratio_i$ / 1 , airway radius $\Radius_i$ / 1, mean air velocity $\Velocity_i$ / 1, Reynolds number $\R_i$ / 1, wall shear stress (Poiseuille) $\ShearStress_i^p$ / 1, wall shear stress (inertia) $\ShearStress_i^I$\\  ($Re_i > 300$) / 1}{,$\frac{1}{2}$,$\sqrt{\frac{1}{2}}$, $\sqrt[3]{\frac{1}{2}}$, $1$,}	
\tkzTabLine{,\searrow,\searrow,\searrow, \searrow,\searrow,\searrow,\searrow,\rightarrow,\nearrow,}
\tkzTabLine{,\nearrow,\nearrow,\nearrow, \rightarrow, \searrow,\searrow,\searrow,\searrow,\searrow,}
\tkzTabLine{,\nearrow,\rightarrow,\searrow, \searrow,\searrow,\searrow,\searrow,\searrow,\searrow,}
\tkzTabLine{,\nearrow,\nearrow,\nearrow, \nearrow,\nearrow,\rightarrow,\searrow,\searrow,\searrow,}
\tkzTabLine{,\nearrow,\nearrow,\nearrow, \rightarrow, \searrow, \searrow,\searrow,\searrow,\searrow,}
\end{tikzpicture}
\caption{Variations of the core quantities involved in our study from one generation $i$ to the next generation $i+1$, depending on the value of the ratio $\Ratio_i$ between the radius of the airways in generation $i+1$ and the radius of the airways in generation $i$. 
The upwards arrows indicate that the quantity is increasing from the mother airway to the daughter airways, the horizontal arrow that it remains constant and the downward arrows that it is decreasing.}
\label{table:Variation}
\end{table*}

\begin{figure*}[p!th]
\centering
\hspace{0.15cm}
A \includegraphics[width=7.5cm]{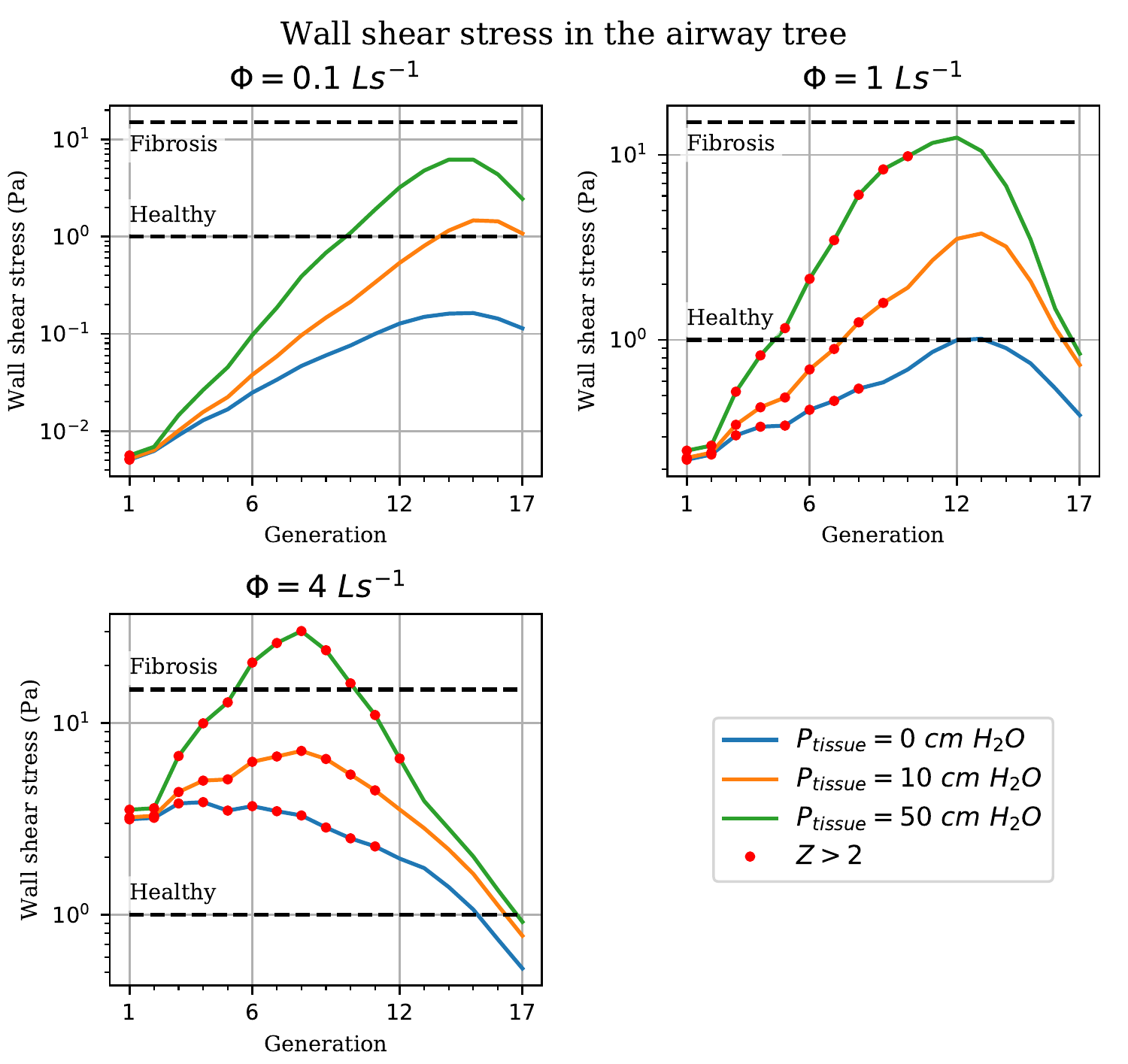}
B \includegraphics[width=7.5cm]{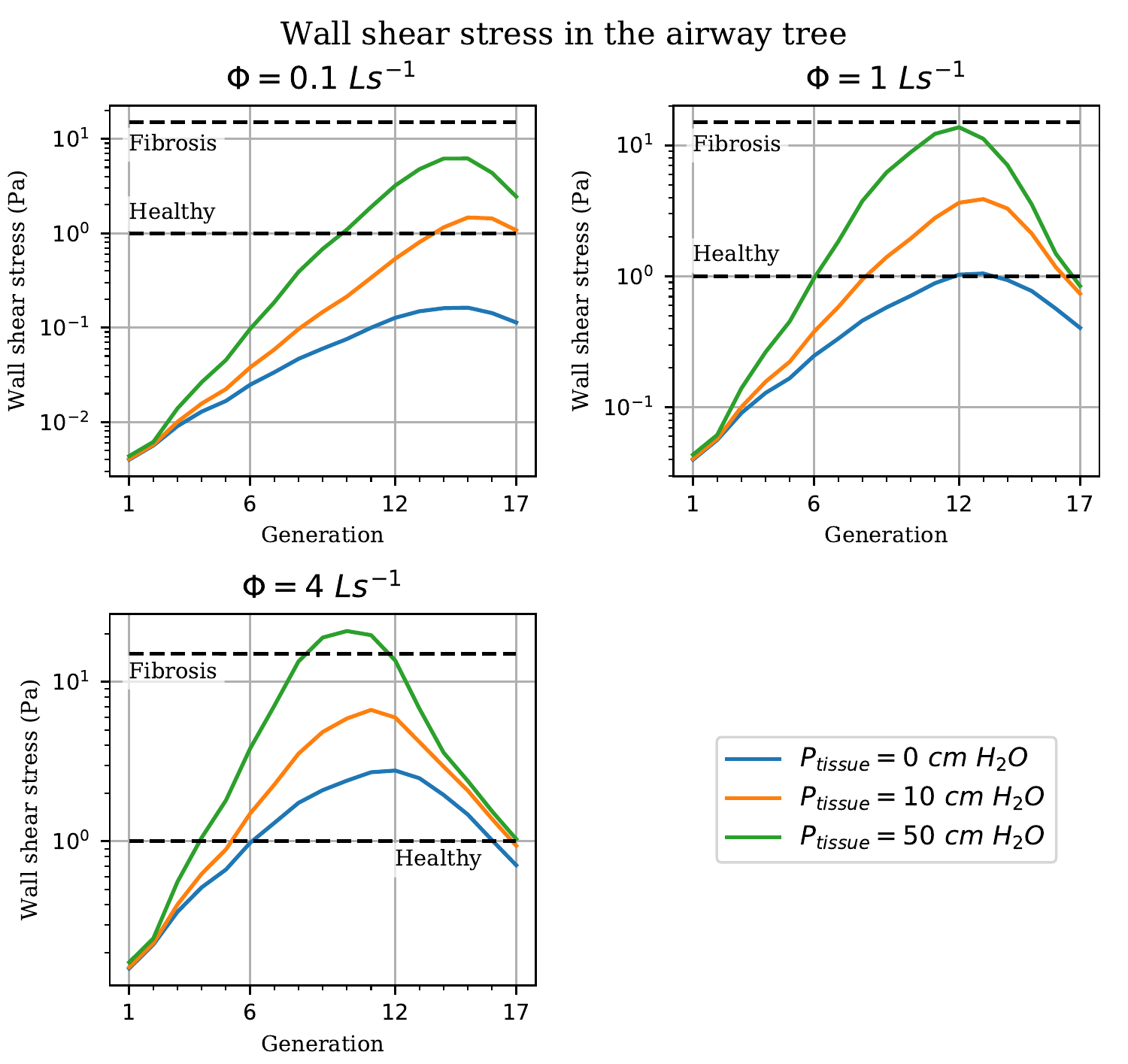}
\newline
\newline
C \includegraphics[width=7.5cm]{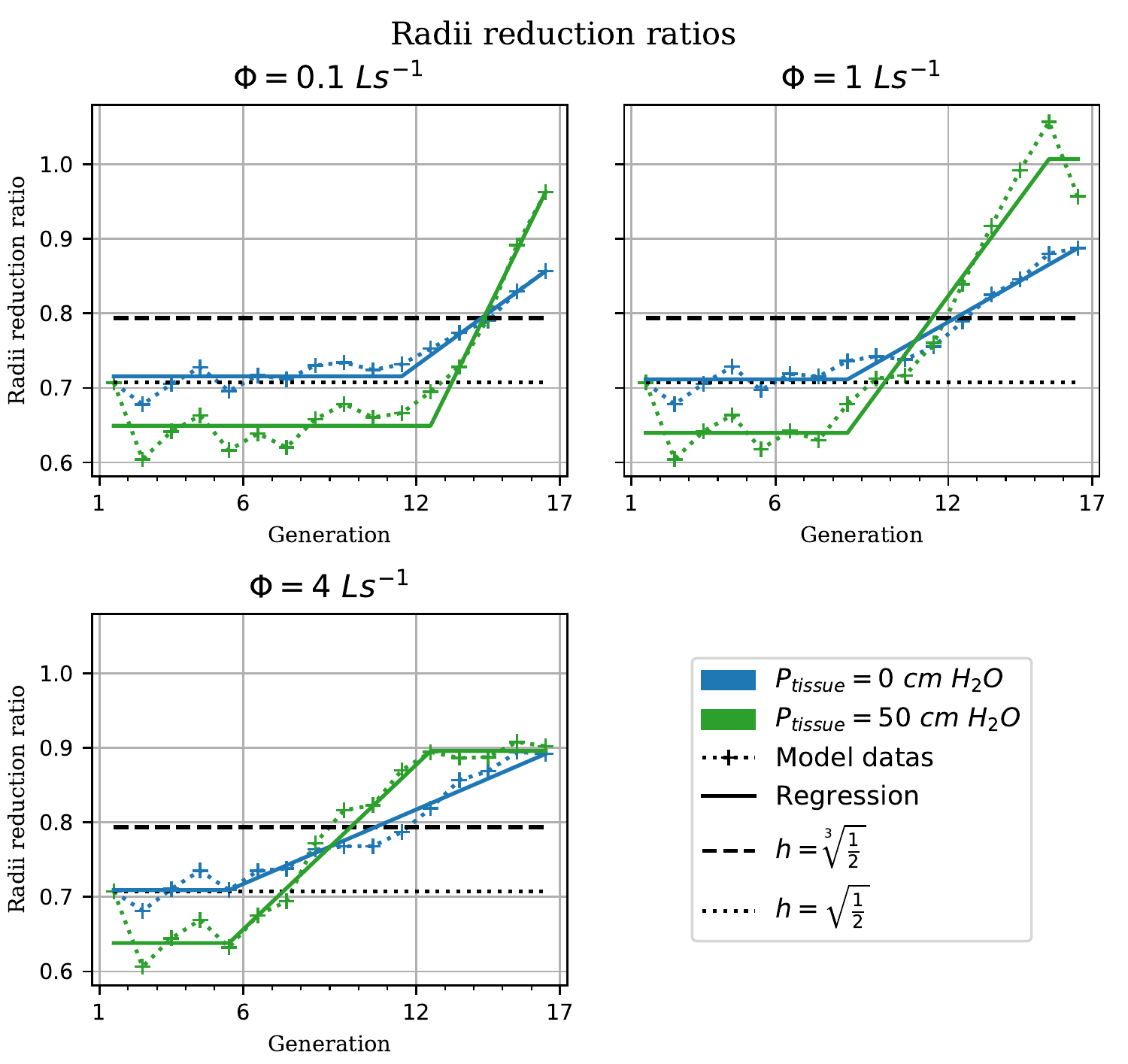}
D \includegraphics[width=7.5cm]{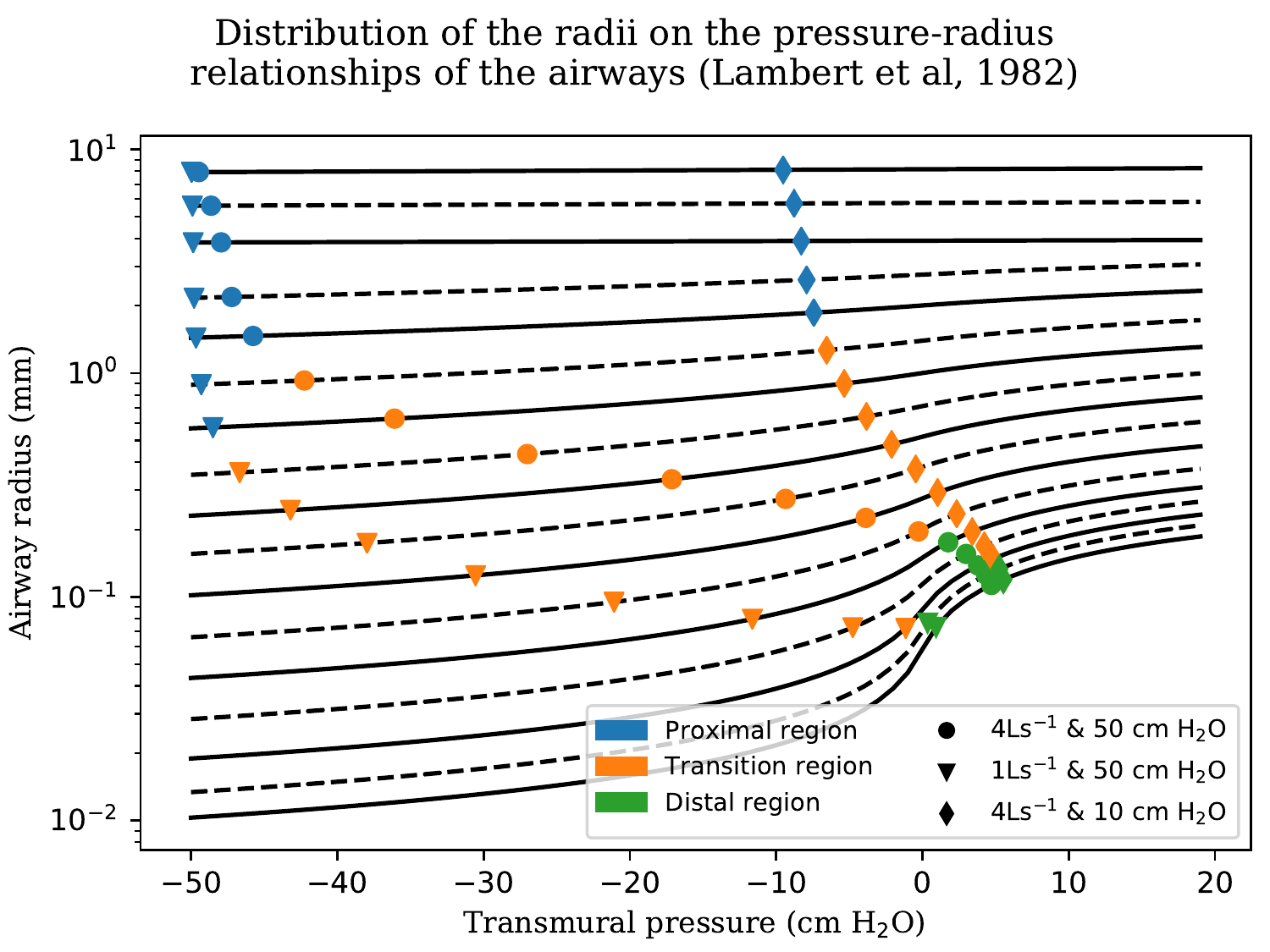}
\caption{
{\bf A, B:} Shear stress distribution along the airway tree as a function of the generation index, for different airflows $\Flow$ and tissue pressures $\ExtPressure$ (A: with inertia, B: without inertia).
The dashed horizontal lines represent an estimation of the the yield stress for healthy mucus ($1$ Pa) and for cystic fibrosis mucus ($15$ Pa).  
The red dots indicates regions where inertia is large, i.e. $\ResistMod \leq 2$ which is equivalent to $\sigma^i > 2 \sigma_P^i$. 
\newline 
{\bf C:} Radius reduction ratios $h_i$ between two successive generations of the airway tree as a function of the generation index $i$, for different airflows $\Flow$ and tissue pressures $\ExtPressure$. 
The ratios exhibit up to three different regimes.
\newline
{\bf D:} Radii of the airways located on their compliance curve as a function of their transmural pressure, for a subset of the cases tested in this study. 
One symbol represents a specific configuration of air flow and pressure.
The horizontal shift between two generations corresponds to the air pressure drop between these geenrations.
}
\label{curves}
\label{fig:ss(01,1,4)}
\label{fig:ratios}
\label{fig:ssm(01,1,4)}
\label{fig:ResultPoiseuille}
\end{figure*}

\subsection{Inertial and viscous critical ratios}
\label{analytic}
Our model predicts that the wall shear stress exhibits a bell distribution with a maximal value somewhere in the airway tree. 
On each side of the maximum, the wall shear stress varies exponentially, increasing in the proximal airway and decreasing in the distal airway, as shown in figure \ref{fig:ss(01,1,4)}A. Actually, we will show that this behavior results from the geometry of the airway tree.

In a bifurcation, the radius $\Radius_{i+1}$ of the daughter airways is a reduction of the radius $\Radius_i$ of the mother airway, $\Radius_{i+1} = \Ratio_i \Radius_{i}$. 
The ratios $\Ratio_i$ allow to derive a scaling for the mean velocity in the airway $\Velocity = \Phi/(\pi R^2)$, with $\Velocity_{i+1} = \frac{1}{2 \Ratio_i^2} \Velocity_i$, and for the Reynolds number $Re = 4 \rho \Phi / (\mu \pi R)$, with $Re_{i+1} = \frac{1}{2 h_i} Re_i$.
For Poiseuille's regime (low flow), we have shown that the change of the wall shear stress through a bifurcation determines the behavior of the wall shear stress at the scale of the airway tree, see \cite{stephano_modeling_2019}.
Indeed, in Poiseuille's regime, the change of the wall shear stress $\ShearStress^p = 4 \mu \Phi / (\pi R^3)$ through a bifurcation scales as $\ShearStress_{i+1}^p = \frac{1}{2 \Ratio_i^3} \ShearStress_i^p$.
Depending on the position of the ratio $\Ratio_i$ relatively to the viscous critical ratio $\hc \simeq 0.793$, the wall shear stress can either increase or decrease through the bifurcation.  
The different scenarios for the wall shear stress and for other physical quantities are given in table \ref{table:Variation}.
For example, in the case of a constant ratio $\Ratio_i = \Ratio$ throughout the airway tree, the radii $\Radius_i$, the velocities $\Velocity_i$ and the mean wall shear stresses $\ShearStress_i^p$ will either grow exponentially, remain constant or decrease exponentially.

In the case studied in this paper, we also account for the inertial effects by deforming the para\-bolic velocity profile of the linear regime into a plug-like velocity profile that depends on the Rey\-nolds number \cite{tawhai_2004}.
The plug like profile induces higher wall velocity gradients than the parabolic profile. 
Consequently, inertial effects increase the pressure drops and the wall shear stresses in the airways.
In the presence of inertia, the shear stress decomposes into $\ShearStress_{i} = \ShearStress_{i}^p / 2 + \ShearStress_{i}^I$ where  $\ShearStress_{i}^I$ reflects the influence of the inertia,
\begin{align*}
&\ShearStress_{i}^I =
\left\{
\begin{array}{ll}
\frac{\ShearStress_{i}^p}{2} = \frac{2 \mu}{\pi} \frac{\Phi_i}{R_i^3} & \text{ if } \R_i \leq 300\\
\frac{8 \rho}{600 \pi^2}\frac{\Phi_i^2}{R_i^4} & \text{ if } \R_i > 300
\end{array}
\right.
\end{align*}
When inertia is dominant ($\R_i >> 1$), the change in wall shear stress between the generations $i$ and $i+1$ depends on the location of $h_i$ relatively to the inertial critical ratio $\hci$, since $\frac{\Phi_{i+1}^2}{R_{i+1}^4} = \left(\frac1{2 h_i^2}\right)^{2} \frac{\Phi_{i}^2}{R_{i}^4}$. 
When inertia is small ($\R \leq 300$ in our model), the change in wall shear stress between the generations $i$ and $i+1$ is that of the Poiseuille regime and, as detailed above, it depends on the location of $h_i$ relatively to the viscous critical ratio $\hc$.

The Reynolds number is multiplied from the generation $i$ to the next by the ratio $\frac1{2h_i}$.
Hence, the influence of inertia on the shear stresses is also changing from one generation to the next with the same rate.
Since $h_i$ is in general larger than $\frac12$, the influence of inertia is decreasing along the generations of the tree and disappears in the distal airways.

\subsection{A bell-shaped distribution of the wall shear stress} 

The bell shape exhibited by the shear stresses in the airway tree results from the interaction of the flow with the geometry of the airway tree.
This interaction is driven by the reduction ratios $h_i$ that correspond to the relative change of the radii between one generation and the next.
If we compute these ratios during idealised manipulations, shown in figure \ref{fig:ratios}B, we can see that the bifurcations along which the shear stress is increasing correspond to the low values of $h_i$ and the bifurcations along which the shear stress is decreasing correspond to the high values of $h_i$.

In the case of a low air flow rate ($0.1$ \LperS), inertia is low and the behavior is controlled by the location of the ratios $h_i$ relatively to the viscous critical ratio $\hc$. 
In this configuration, the generation at which the maximum of the shear stress is reached is exactly where the transition $h_i = \hc$ occurs, see figure \ref{fig:ratios}A and B ($0.1$ \LperS), table \ref{table:Variation} and \cite{stephano_modeling_2019}.
In the presence of inertia, the shear stress in the proximal airways is increased relatively to the Poiseuille regime, see figure \ref{fig:ss(01,1,4)}A and C.
The shape of the wall shear stress distribution and the location of its maximum result from the relative contribution and evolution along the generations of the two terms in the decomposition of the shear stress, $\ShearStress_{i} = \ShearStress_{i}^p / 2 + \ShearStress_{i}^I$.
For the large flows, typically $4$ \LperS, the shear stress in the proximal airways is dominated by inertia, i.e. $\ShearStress_{i}^I >> \ShearStress_{i}^p$.
Then, the maximum of the shear stress is located where $h_i$ exceeds the inertial critical ratio $\hci \simeq 0.707$, see figures \ref{fig:ratios}A and B ($4$ \LperS) and table \ref{table:Variation}.
\section{Role of the fluid dynamics}

\subsection{Three regions, three behaviors}
The distribution of the ratios $h_i$ along the generations of the tree can be decomposed into three regions, see figure \ref{fig:ratios}B: a proximal region, where the diameters reduction ratios are low; A transitional region, where the diameters reduction ratios increase; A distal region, not present at low flow rates, where the diameters reduction ratios are high.
The three regions arise from different resolutions of a trade-off between the air pressure in the airways and the air pressure drops in the airways bifurcations.

{\bf Proximal region.} The air pressure drops in the large airways are low, even in the presence of inertia.
Hence, the airway diameters are mainly affected by the tissue pressure. 
As the daughter airways in a bifurcation are more compliant than the mother airway, they are more constricted. 
Consequently, their radii reduction ratios are smaller than the ratios at functional residual volume (FRC) shown in table \ref{ratioLambert}.
Hence, the ratios tend to be close to the inertial critical ratio $\hci \simeq 0.707$ and smaller than the viscous critical ratio $\hc \simeq 0.793$.
Consequently, in the proximal region, the shear stress is in general increasing from one generation to the next.

{\bf Transitional region.} 
The diameters of the airways are decreasing along the generations, see figure \ref{curves}D.
Hence, the pressure drops are increasing, at least in the first generations of the transitional region.
The change of the transmural pressure through a bifurcation starts to compensate the lower compliance of the daughter airways.
The daughter airways are actually less compressed relatively to the mother airway than in the proximal region.
This results in higher ratios $h_i$.
This effect becomes stronger and stronger along the generations, and the ratios increase steadily.
Eventually, they exceed $\hci$ and $\hc$, and
the maximum of the shear stress and of the pressure drop is reached.

{\bf Distal region.} 
In the distal region, the transmural pressure is slightly positive, at least in the cases that we tested. 
The small transmural pressures in the distal region corresponds to near maximal compliances values, as they correspond to large slopes zones on the pressure--section curves shown on figure \ref{curves}D.
Moreover, since the smaller airways have larger compliances and higher inner air pressures, they are, relatively, more dilated than the larger ones.
As a consequence, the reduction ratios are relatively high in the distal region, reaching values near $0.9$ or higher. 

\subsection{Dependance on air flow rates and tissue pressures} 

The distribution and amplitude of the shear stresses in the airway tree depend on the flow rate $\Phi$ and on the tissue pressure $\ExtPressure$, more particularly the position and amplitude of its maximum, see figure \ref{curves}. 
The shear stress depends on the ratio between the air flow and the cube of the radius of the airway, $\sigma = \frac{\alpha + 2}{4} \frac{8 \mu}{\pi} \frac{|\Phi|}{R^3}$.
An increase of the air flow rate induces higher pressures in the airways, which are more dilated.
The variation of the shear stress is the result of the balance between the air flow rate increase and the increase of the cube of the airways' radii.
Since our model predicts that the amplitude of the maximal shear stress increases with the air flow rate, this suggests that the increase of the air flow rate might be dominant over the increase of the airways' radii, at least in the range of the parameters tested.  
In this configuration, pressure drops in the airways are larger because of the higher flows and the air pressure in the airways increases more quickly along the generations. 
Hence, both the proximal and the transitional regions are covering less generations.
The location of the maximum of the shear stress is then shifted toward the proximal generations of the airway tree.

When the tissue pressure increases at constant air flow rate, the amplitude of the shear stresses also increases, see figure \ref{curves}.
This phenomenon is due to the resulting decrease of the radii of the airways.
Hence, the pressure drops between two successive generations are increased.
In the proximal region, the reduction ratios are smaller due to the increase of the compliance with the generation index.
However, the inner air pressures needed to increase the ratios are also higher.
As long as these inner air pressures remain low, the transmural pressures change very little.
As a consequence, the range of generations covered by the proximal region remains the same.
The two other regions are however affected by the larger pressure drops.
In the transitional region, the reduction ratios increase more quickly, hence reducing the range of generations covered by the region.
In the distal region, the air pressure is high as it results from the addition of all the pressure drops.
The airways are then closer to their maximal dilation, and for high tissue pressures the reduction ratios are near $0.9$, see table \ref{ratioLambert}.

Our model predicts that the location in the airway tree of the maximum of the shear stress depends only slightly on the tissue pressure. 
This phenomenon results from the variation of the reduction ratios in the transitional region.
In that region, if the tissue pressure is increased, the reduction ratios are increasing from values that are lower, as the higher tissue pressure decreases the ratios in the proximal region.
Nevertheless, in the transitional region, the increase of the ratios from one generation to the next occurs at a higher pace, due to the higher pressure drops.
This results in an increase in reduction ratios that compensates the lower values in the proximal region.
This allows to reach the critical value of the ratio, which corresponds to the maximum of the shear stress, in a generation index close to that of a lower tissue pressure.

\cred
\section{Model limitations}

Our predictions should be considered as qualitative only, since our model is based on several simplification hypotheses.
First, both the fluid dynamics and the airways deformation \cite{lambert_1982} are considered stationary.
Moreover, we assume that the air velocity profiles in the airways are not affected by the bifurcations and are unchanged along the axis of the airways.
Turbulence was neglected as it can scarcely be established in the airways, with the notable exception of the trachea \cite{lin_characteristics_2007}. 
Actually, in most of the airways, the air flow reaches the bifurcation before the turbulence could really develop and have some influence on the pressure drops. 
Also, neglecting turbulence allows to get a more tractable model. 
Nevertheless, a more precise evaluation of the role of turbulence will be made in future evolutions of our model.
In term of geometry, the airway tree mimicking the lung in our model has symmetric bifurcations, unlike the lung where asymmetric branching is known to affect the fluid dynamics \cite{mauroy_2004, buess_2019}.
Also, our airways remain perfectly cylindrical whatever their transmural pressure. 
The typical star shape of the lumen of a constricted airway is not accounted for~\cite{mauroy_role_2015}.
A total collapse of the airways is also not possible in our model since the air flow rate is constrained.
The computation of the transmural pressure is an approximation, as we assume an homogeneous tissue pressure all throughout the lung.
Finally, we neglect the pressure drops induced by the bifurcations and by the nasopharyngeal or buccopharyngeal pathway.
Actually, only a full 3D analysis of the different regimes could bring adequate estimations of these pressure drops \cite{paz_glottis_2017, longest_use_2019}. 
Such a study is out of the scope of this paper but should be performed in the near future in order to reach better estimations of the pressure drops.

Nevertheless, our model includes the main physical and physiological phenomena affecting the wall shear stress in the lung: the airway tree structure with deformable airways and the resulting fluid dynamics, including some inertial effects.
Hence, the predictions of our model should highlight characteristic behaviors for the wall shear stress distributions in airway trees such as the lung. 
\cb

\section{Discussion} 

\cred
Our work predicts the distribution of the wall shear stress induced by an air flow exiting an airway tree surrounded by an external pressure.
The wall shear stress exhibits a bell-shaped distribution a\-long the generations of the tree. 
This shape results from the complex, non-linear interactions between the fluid dynamics, the geometry of the tree and the compliance of the airways.
Our analyses allow to uncover the dynamics of these interactions.
We show that the change in diameter of the airways through a bifurcation determines if the wall shear stress increases or decreases through that bifurcation.
We derived analytical criteria that characterize specific regimes. 
They are based on the ratio $h_i$ between the diameters of the daughter airways and the mother airway, see table \ref{table:Variation}. 
In Poiseuille's regime, the wall shear stress increases through a bifurcation only if $h_i$ is larger than the viscous critical ratio $\left(\frac12\right)^{\frac13} \simeq 0.793$ and decreases otherwise.
In regimes where inertia is dominant, the inertial critical ratio is lower and equal to $\left(\frac12\right)^{\frac12} \simeq 0.707$.
When both viscous and inertial influences are of the comparable order of magnitude, then the corresponding critical ratio stands between the viscous and the inertial ones.

Our results give insights about the air--mucus interactions at expiration, not only during normal ventilation but also during chest physiotherapy.
Chest physiotherapy uses the air wall shear stress to help the draining of the mucus from the bronchial tree.
But in order for mucus to be motioned, the mucus yield stress has to be overcome. 
Our model suggests that a maximum of the wall shear stress can be, to some extent, localized in the airway tree by tuning the tissue pressure and the air flow rate.
The existence of an adjustable maximum of wall shear stress by accessible physical parameters is of great importance for validating scientifically the technics used by chest physiotherapists. 
Our results suggest that the chest physiotherapists might intuitively manipulate the location and the amplitude of the maximum of the wall shear stress in order to overcome the yield stress of the mucus in specific regions of the lung.

Our results and analyzes can be obtained only because our model is a "minimal model". 
Minimal is meant here in the sense that the model includes only the core physical properties of the lung's biomechanics: the tree structure, the deformation of the airways, the air fluid dynamics with inertia and an external tissue pressure.
The first benefit in using a minimal model is to catch and to interpret more easily the dynamics of the system. 
The effects of each physical process can be isolated and confronted to the others.
The second benefit in using a minimal model is that the model is numerically tractable.
This allows the exploration of many scenarios as its computation time is cheap.
Tractable models such as ours are crucial for analyzing future models with higher complexity.

Future evolutions of our model will include a more realistic tree structure that accounts for the asymmetry of the bifurcations \cite{tawhai_2004}.
Moreover, we plan to mimic both the entering (inspiration) and the exiting (expiration) airflows.
We will also evaluate the role of the turbulence in the airways with large Reynolds numbers.
Finally, in the lung, the tissue pressure and the amount of airflow are related. 
In order to link these two physical quantities, we plan to couple an idealised 0D model of the lung's tissue mechanics to the model presented in this paper.
This will allow to mimic lung's configurations that reflect better the physiology of the lung, most particularly during chest physiotherapy.
\cb

\section{Conclusion}

This work aims to improve our understanding of how the wall shear stresses are distributed during expiration in an airway tree with deformable airways.
The wall shear stress has a bell-shaped distribution along the generations of the tree. 
This shape results from the interactions between the geometry of the tree, the compliance of the airways and the air fluid dynamics.
The proximal part of the bell and the location of the maximum are dependent on the presence of inertia.
Our analyses show that the properties of the distribution of the wall shear stress, and more particularly the location of the maximum, are affected by the air flow rate and the tissue pressure.
In the limit of our model, these results could allow to do physics-based analyses of several commonly used chest physiotherapy technics.
Moreover, our model suggests that the control of these two quantities during chest physiotherapy manipulations might allow to focus the draining of the mucus on a specific region in the lung.
Hence, our work might help to understand, or possibly improve, the current mucus draining strategies.

\cred
\section*{Acknowledgement}

The authors would like to thank the physiotherapists Jean-Claude Jeulin, Christian Fausser and Dominique Pelca for fruitful discussions.

This work has been supported by the Centre National de la Recherche Scientifique (CNRS) and by the Agence National de la Recherche in the frame of the project VirtualChest (ANR-16-CE19-0014) and of the IDEX UCA JEDI (ANR-15-IDEX-01).

\section*{Data Availability Statement}

The data that support the findings of this study are available from the corresponding author upon reasonable request.
\cb

\bibliographystyle{apsrev4-1}

\newpage

\onecolumn

\appendix

\begin{center}
{\huge \bf Appendix}
\end{center}

\section{Derivation of the relationship between the wall shear stress and the pressure drop}
\label{sigmaq}
\begin{figure}[h!]
\centering
\includegraphics[width=6cm]{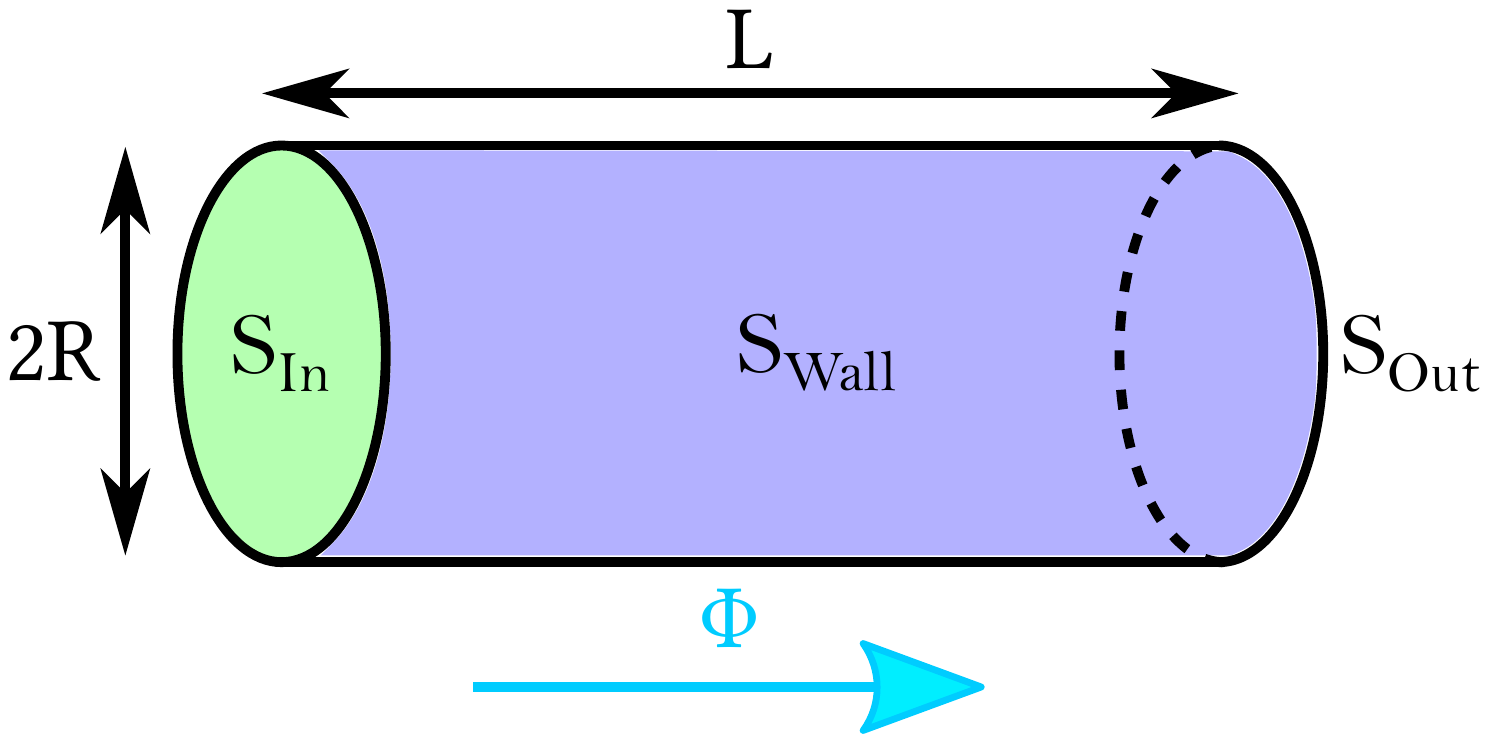}
\caption{Schematics of a cylindrical airway: the cylinder has a length $\Length$ and a diameter $2\Radius$. The inlet is denoted $\Inlet$, the outlet $\Outlet$ and the wall $\Wall$. The air flow through the airway is denoted $\Phi$.}
\label{fig:TubeScheme}
\end{figure}

We consider a cylindrical airway $\Bronchus$ with a radius $\Radius$ and a length $\Length$.
We use cylindrical coordinates and the axis of the airway is aligned with the axis $\hat{Z}$. 
The airway is delimited by three surfaces: the airway wall $\Wall$ where the air velocity is assumed to be zero (non slip boundary conditions), the normal vector to $\Wall$ is denoted $\hat{r}$; the airway inlet $\Inlet$, where the air flow rate is $\Flow = \iint_{\Inlet} \Velocity_z d\Surface$ with $\Velocity_z$ the component of the air velocity along the axis of the cylinder, the normal vector to $\Inlet$ is $-\hat{z}$; the airway outlet $\Outlet$ where the air flow rate equals the inlet flow rate, $\Flow = \iint_{\Outlet} \Velocity_z d\Surface$, the normal vector to $\Outlet$ is $\hat{z}$. 
The total surface of the airway is denoted $S_{\Bronchus} = \Inlet \cup \Outlet \cup \Wall$. A schematics of the airway is shown in figure \ref{fig:TubeScheme}.

The fluid dynamics of the air in the airway is driven by the Navier-Stokes equations that relates the pressure $p$ and the velocity $\bar{\Velocity}$ of the air,
\begin{equation}
\left\lbrace \begin{array}{l}
\Density \partial_t \bar{\Velocity} + \Density \bar{\nabla}\cdot\left(\bar{\Velocity}\otimes \bar{\Velocity}\right) = - \bar{\nabla} \Pressure + \Viscosity \Delta \bar{\Velocity}\\
\bar{\nabla}\cdot\bar{\Velocity} = 0
\end{array}\right.
\label{NavierStokes}
\end{equation}

The shear stress is the consequence of the friction between the air layers and between the air and the airway's wall. The friction is related to the viscosity of the air, i.e. the term $\Viscosity \Delta \bar{\Velocity}$ in the Navier-Stokes equations. Due to the stationary hypothesis, we have $\partial \bar{\Velocity} / \partial t = 0$.

We denote $\Pressure_{\rm{In}}$ and $\Pressure_{\rm{Out}}$ the mean pressures at the inlet and the outlet: $\Pressure_{*} = (\pi\Radius^2)^{-1} \iint_{\Surface_*} \Pressure\ d\Surface$ with $*=\rm{In}$ or $\rm{Out}$. We define the wall shear stress $\ShearStress$ as the mean norm of the friction constraint on the airway wall:
$\ShearStress=\Viscosity(2\pi\Radius\Length)^{-1}\iint_{\Wall} || \partial \bar{\Velocity} / \partial r || d\Surface $.

Then, by integrating the Navier-Stokes equations (\ref{NavierStokes}) on the volume of the airway $\Volume_{\Bronchus}$, using the Stokes theorem and projecting the result on the $\hat{Z}$ axis, we have
\begin{equation}
\Density \oiint_{\Surface_\Bronchus} \bar{\dd\Surface} \cdot \left(\bar{\Velocity} \otimes \bar{\Velocity}\right) \cdot \hat{z} = \iint_{\Outlet} \dd\Surface\ \Density \Velocity_z^2 - \iint_{\Inlet} \dd\Surface\  \Density \Velocity_z^2
\label{eq:BronchusScale}
\end{equation}

The term $\Density \Velocity_z^2$ is the local kinetic energy of the air in the direction of the axis of the airway. Due to the assumption that the velocity profile is invariant along the airway axis $\hat{Z}$, the axial kinetic energy is conserved all along the airway and the previous term is zero.

The term resulting from the $\bar{\nabla} p$ in the Navier-Stokes equations can be expressed using the airway pressure drop $\PressureDrop=\Pressure_{\rm{In}} - \Pressure_{\rm{Out}}$:
\begin{align*}
\oiint_{\Surface_\Bronchus} \bar{\dd\Surface}\cdot \hat{z}\ \Pressure &= \iint_{\Inlet} \dd\Surface\ \Pressure - \iint_{\Outlet} \dd\Surface \Pressure\\
&= \pi \Radius^2 \PressureDrop
\end{align*}

The viscous term $\Viscosity \Delta \bar{\Velocity}$ becomes:
\begin{equation*}
\Viscosity \oiint_{\Surface_\Bronchus} \bar{\dd\Surface} \cdot \bar{\nabla} \otimes \bar{\Velocity} \cdot \hat{z} =\Viscosity\!\! \iint_{\Outlet}\!\! \dd\Surface\ \frac{\partial\Velocity_z}{\partial z}  - \Viscosity\iint_{\Inlet} \dd\Surface\ \frac{\partial\Velocity_z}{\partial z} + \Viscosity\!\!\! \iint_{\Wall}\!\! \dd\Surface\ \frac{\partial\Velocity_z}{\partial r}
\end{equation*}

Under the assumption that the air velocity is invariant along the axis of the airway, the quantity $\partial\bar{\Velocity}/\partial z$ is zero, hence
\begin{equation*}
\Viscosity \oiint_{\Surface_\Bronchus} \bar{\dd\Surface} \cdot \bar{\nabla} \otimes \bar{\Velocity} \cdot \hat{z} = \Viscosity\!\!\!\iint_{\Wall}\!\! \dd\Surface\ \frac{\partial\Velocity_z}{\partial r}
\end{equation*}

Finally, the hypotheses of axial and axisymmetric air velocities that are unchanging along $\hat{Z}$ allow to write $||\Viscosity \partial\bar{\Velocity}/\partial r|| = |\Viscosity \partial\Velocity_z/\partial r|$. Using our definition for the mean wall shear stress $\ShearStress$ leads to
\begin{align*}
\Viscosity \oiint_{\Surface_\Bronchus} \bar{\dd\Surface} \cdot \bar{\nabla} \otimes \bar{\Velocity} \cdot \hat{z} &= \pm \Viscosity\!\!\!\iint_{\Wall}\!\! \dd\Surface\ |\frac{\partial\Velocity_z}{\partial r}|\\
&=\pm \Viscosity\!\!\!\iint_{\Wall}\!\! \dd\Surface\ || \frac{\partial\bar{\Velocity}}{\partial r}||\\
&=\pm 2\pi\Radius\Length\ShearStress
\end{align*}

Then regrouping all the previous transformed terms as in equations \ref{NavierStokes}, we obtain:
\begin{equation}
\label{eq:shear2pressure}
\ShearStress = |\PressureDrop \frac{\Radius}{2\Length}|
\end{equation}
The wall shear stress is directed in the same direction as the airflow and in the opposite direction of the pressure drop. The signed mean shear stress relationship is then $\ShearStress = -\PressureDrop \frac{\Radius}{2\Length}$. 

\section{Solving the model's equations}

\label{system}

The section \ref{model} describes the model used in this work. 
The system of equations that needs to be solved is based on one sub-system for each generation $i$ of the airway tree that writes
\begin{equation}
\left\lbrace\begin{array}{lcl}
\PressureDrop_i &=& \frac{8\Viscosity}{\pi} \ResistMod_i \Length_i\Flow_i \Radius_i^{-4}\\
\ResistMod_i &=& \max\left(1,\frac12 + \frac{\Reynolds_i}{600}\right)\\
\Reynolds_i &=& \frac{4\Density \Flow_i}{\Viscosity\pi\Radius_i}\\
\Radius_i &=& R_{max,i} \sqrt{\frac{\NormSection_i}{\pi}}\\
\NormSection_i & =& \Lambert_{,i}\left(\TransPressure_i  \right)\\
\TransPressure_i &=& \sum_{k=0}^{i-1}\PressureDrop_k + \frac{\PressureDrop_i}{2} - P_{tissue}\\
\ShearStress_i &=& \frac{\PressureDrop_i\Radius_i}{2\Length_i}\\
\Phi_i &=& \Phi_{i-1} / 2
\end{array}\right.
\label{eq:system}
\end{equation}

The pressure drop $\PressureDrop_i$ in the airways in generation $i$ depends on their radius $\Radius_i$ which depends in turn on the pressure drop $\PressureDrop_i$. 
The shear stress $\ShearStress_i$ can be computed only when the pressure drop $\PressureDrop_i$ is known. So, solving the system is equivalent to find a fixed point $\PressureDrop_i$ of the previous system of equations (\ref{eq:system}). 
Hence, the system can be reformulated into finding the fixed point $q_i$ of an application $F_i$ that computes a pressure drop $F_i\left(\PressureDrop\right)$ from a given pressure drop $\PressureDrop$ in generation $i$. 

For any generation index $i$, the pressure drops $\PressureDrop_k$ for $k<i$ are the fixed points of the systems associated to the generations $k < i$. Hence, we can compute all the fixed points of the tree-wide system by starting with the generation $0$ and progressing downward the tree, down to the final one. A Newton algorithm is ran to find the zero of the application $F_i(\PressureDrop) - \PressureDrop$ to determine the fixed point $\PressureDrop_i$. In the Supplementary Materials (\ref{appendix}), we show that the system has only one solution, hence ensuring that the convergence of the Newton algorithm leads to the correct pressure drops distribution. In the Supplementary Materials (\ref{numerics}), details about the numerical method are given. 

\section{Existence and uniqueness of the solution of the model equations}
\label{properties}
\label{appendix}

In order to be sure that our algorithm converges towards the correct physical solution, we need to highlight the existence and unicity of the solution.

The first step is to study the behavior of the quantity $F_i\left(\PressureDrop\right) - \PressureDrop$.
The derivative of $F_i\left(\PressureDrop\right)$ relatively to $q$ writes
\begin{equation}
\left(\partial_{\PressureDrop} F_i\right)(\PressureDrop) = 
\left(\partial_{\Radius_i} F_i \cdot
\partial_{\NormSection_i} \Radius_i \cdot
\partial_{\TransPressure_i} \NormSection_i \cdot
\partial_{\PressureDrop}\TransPressure_i\right)(\PressureDrop)
\end{equation}
$F_i(\PressureDrop)$ is a function of $\Radius_i$ that can be expressed as $F_i(\PressureDrop) = a \Radius^{-4}_i + b \Radius^{-5}_i$, with $a$ and $b$ data that depends on the flow rate $\Flow_i$ and on the hydrodynamic resistance. 
Since, we know that $a$ and $b$ are always positive, $\partial_{\Radius_i} F_i(\PressureDrop)$ is always negative:
\begin{align}
\left(\partial_{\Radius_i} F_i\right)(\PressureDrop) &= - 4 a\ \Radius_i^{-5} - 5 b\ \Radius_i^{-6} &<0
\end{align}
Now, for $\partial_{\NormSection_i} \Radius_i$, we have:
\begin{equation}
\partial_{\NormSection_i} \Radius_i = \frac{\Radius_{max,i}}{2\sqrt{\pi \NormSection_i}} >0
\end{equation}
Next, for $\partial_{\TransPressure_i} \NormSection_i$, we have:
\begin{align}
\partial_{\TransPressure_i} \NormSection_i &\stackrel[\TransPressure_i\le 0]{}{=}n_i^-\frac{\NormSection_{0,i}}{P_i^-}\left(1-\frac{\TransPressure_i}{P_i^-}\right)^{-n_i^--1} & >0\\
&\stackrel[\TransPressure_i\ge 0]{}{=}n_i^+\frac{1-\NormSection_{0,i}}{\TransPressure_i^+}\left(1+\frac{P_i}{P_i^+}\right)^{-n_i^+-1} &>0
\end{align}
Finally, for $\partial_{\PressureDrop_i}\TransPressure_i$, we have:
\begin{equation}
\partial_{\PressureDrop}\TransPressure_i = \frac12 > 0
\end{equation}
This leads to 
\begin{equation}
\forall \ q \in \mathbb{R} \ \left(\partial_{\PressureDrop} F_i\right)(\PressureDrop) -1 < -1
\end{equation}
So, $F_i(\PressureDrop) - \PressureDrop$ is strictly decreasing and $\lim_{q \rightarrow +\infty} F_i(q) - q = -\infty$.\\

The next step is to show that the application $F_i(\PressureDrop) - \PressureDrop$ is bijective from $\mathbb{R}$ onto $\mathbb{R}$, i.e. that it can reach any real value with a unique value of $q$.
The model for airway compliance makes the values of the airway radius $R_i$ range unequivocally from $0$, at a limit transmural pressure $P_i = -\infty$, to $R_{max,i}$, at a limit transmural pressure $P_i = +\infty$.
The dependance of $R_i$ on $q$ is identical, since the transmural pressure $P_i$ is linearly and positively dependant on the pressure drop $q$.  
Hence, the application $F_i(q)$ that depends on $1/R_i^4$ ranges from $+\infty$ when $q$ goes to $-\infty$ to a limit value $q_{inf,i}$ when $q$ goes to $+\infty$.
We can conclude that the application $F_i(q) - q$ ranges from $+\infty$ when $q$ goes to $-\infty$ to $-\infty$ when $q$ goes to $+\infty$.
Finally, adding the fact that the application $F_i(q) - q$ is strictly decreasing, we can conclude that $F_i(q) - q$ is bijective from $\mathbb{R}$ to $\mathbb{R}$

Hence, there is one, and only one, solution $F_i(\PressureDrop) - \PressureDrop = 0$ for the equations (\ref{eq:system}).

\section{Numerical method for solving the system}

\label{numerics}

For each generation $i$, we start by testing if the inertia is affecting the flow profile. 
To get this information, we use the fact that  $F_i(\PressureDrop)$ is decreasing.
A simple graphical analysis shows that, for any pressure drop $q_a$, the solution of $F_i(\PressureDrop) = \PressureDrop$ lies between the values of $\PressureDrop_a$ and $F_i(\PressureDrop_a)$. 
We define the pressure drop $\PressureDrop|_{\R=300}$ that corresponds to a Reynolds number in the airway equal to $300$.
The pressure drop $\PressureDrop|_{\R=300}$ at $\R = 300$ is computed with the Reynolds number formula to compute the corresponding airway radius $R_i|_{\R = 300}$. 
The radius $R_i|_{\R = 300}$ is then injected into the formula that links the pressure drop to the air flow, see equation (\ref{eq:system}).  
As the solution $\PressureDrop^{\star}$ of $F_i(\PressureDrop) - \PressureDrop = 0$ stands between $\PressureDrop|_{\R=300}$ and $F_i(\PressureDrop|_{\R=300})$, two cases are possible.
The first case corresponds to $\PressureDrop^{\star} \leq \PressureDrop|_{\R=300}$. 
As the solution of the equation has a pressure drop which is smaller than $\PressureDrop|_{\R=300}$, the transmural pressure is lower and the airway is more constricted. 
A smaller airway radius induces an airway Reynolds number larger than $300$.
Consequently, we have to account for inertia in our model.
The second case is the opposite and a similar reasoning leads to an airway Reynolds number for the solution of the equations that is smaller than $300$.
The air fluid dynamics in our model is then in Poiseuille regime.

Finally, we use a Newton algorithm to search for the solution of the equation $F_i(\PressureDrop) - \PressureDrop = 0$ for each generation $i$ from the root of the tree to the leaves of the tree. 
Starting with the initial condition $q_{0,i} = 0$, the numerical scheme in the generation $i$ writes:
\begin{equation}
\PressureDrop_{j+1,i} \leftarrow \PressureDrop_{j,i} - \frac{F_i\left(\PressureDrop_{j,i}\right) - \PressureDrop_{j,i}} {\left(\partial_{\PressureDrop} F_i\right) \left(\PressureDrop_{j,i}\right) -1}
\label{eq:NewtonUpdate}
\end{equation}
Here the quantity $\PressureDrop_{j,i}$ is the value computed for the pressure drop $\PressureDrop$ at the j\ith step of the iterative algorithm.

\end{document}